\documentclass{iopart}
\usepackage[T1]{fontenc}
\usepackage{graphicx}
\usepackage{url,graphics,epsfig}
\usepackage{amssymb}

\begin{document}

\jl{2}
%
%+++++++++++++++++++++++++++++++++++++++++++++++++++++++++++++++++++++++++++++
%
%  Macro definitions
%
%+++++++++++++++++++++++++++++++++++++++++++++++++++++++++++++++++++++++++++++
\def\etal{{\it et al~}}
%+++++++++++++++++++++++++++++++++++++++++++++++++++++++++++++++++++++++++++++
%
%  End of Macro definitions
%
%+++++++++++++++++++++++++++++++++++++++++++++++++++++++++++++++++++++++++++++
%
%+++++++++++++++++++++++++++++++++++++++++++++++++++++++++++++++++++++++++++++
%
% Title of the paper
%
%+++++++++++++++++++++++++++++++++++++++++++++++++++++++++++++++++++++++++++++
%
\setlength{\arraycolsep}{2.5pt}             % use this for journal style

\title[K-shell photoionization of Li-like carbon ions]{K-shell photoionization of ground-state Li-like carbon ions [C$^{3+}$]:
       experiment, theory and comparison with time-reversed photorecombination}

\author{A M{\"u}ller$^1$, S Schippers$^1$, R A Phaneuf$^2$, S W J Scully$^{2,3}$,  A Aguilar$^{2,4}$, 
              A M Covington$^2$, I \'{A}lvarez$^5$,  C Cisneros$^5$, E D Emmons$^2$,
               M F Gharaibeh$^2\footnote[1]{Present address: Department of Physics, Jordan University of Science and Technology,
                                                                           Irbid, 22110, Jordan}$,
              G Hinojosa$^5$, A S Schlachter$^4$ 
              and B M McLaughlin$^{3,6}\footnote[2]{Corresponding author, e-mail: b.mclaughlin@qub.ac.uk}$}

\address{$^1$Institut f{\"u}r Atom-und Molek{\"u}lphysik, 
             Justus-Liebig-Universit{\"a}t Giessen, 35392 Giessen, Germany}

\address{$^2$Department of Physics, University of Nevada, 
                          Reno, Nevada 89557, USA}

\address{$^3$School of Mathematics and Physics, The David Bates Building, 7 College Park,
                         Queen's University Belfast, Belfast BT7 1NN, UK}

\address{$^4$Lawrence Berkeley National Laboratory, Berkeley, California 94720, USA }

\address{$^5$Centro de Ciencias F\'isicas, Universidad Nacional Aut\'onoma de M\'exico, 
                           Apartado Postal 6-96, Cuernavaca 62131, Mexico}

\address{$^6$Institute of Theoretical Atomic and Molecular Physics,
                          Harvard Smithsonian Center for Astrophysics, MS-14,
                          Cambridge, Massachusetts 02138, USA}
      
%+++++++++++++++++++++++++++++++++++++++++++++++++++++++++++++++++++++++++++++
%
%              Abstract
%
%+++++++++++++++++++++++++++++++++++++++++++++++++++++++++++++++++++++++++++++
\begin{abstract}
Absolute cross sections for the K-shell photoionization of ground-state
Li-like carbon [C$^{3+}$(1s$^2$2s~$^2$S)] ions were measured by
employing the ion-photon merged-beams technique at the
Advanced Light Source. The energy ranges 299.8--300.15 eV,
303.29--303.58 eV and 335.61--337.57 eV of the
[1s(2s2p)$^3$P]$^2$P, [1s(2s2p)$^1$P]$^2$P and
[(1s2s)$^3$S~3p]$^2$P resonances, respectively, were
investigated using resolving powers of up to 6000. The autoionization linewidth
of the [1s(2s2p)$^1$P]$^2$P resonance was measured to be
$27 \pm 5$~meV and compares favourably with a theoretical result 
of 26 meV obtained from the intermediate coupling R-Matrix method. 
The present photoionization cross section results are compared with the
outcome from photorecombination measurements by employing
the principle of detailed balance. 
\end{abstract} 
%\submitto{\jpb}
%\maketitle
%\submitted

\vspace{1.0cm}
\begin{flushleft}
Short title: K-shell photoionization of C$^{3+}$ ions\\
%\vspace{1cm} Finalized draft for J. Phys. B: At. Mol. \& Opt. Phys: \today
Submitted to  J. Phys. B: At. Mol. \& Opt. Phys: \today
\end{flushleft}

\maketitle
%
%++++++++++++++++++++++++++++++++++++++++++++++++++++++++++++++++++++++++++++
%
%      Text of paper follows
%
%++++++++++++++++++++++++++++++++++++++++++++++++++++++++++++++++++++++++++++

\section{Introduction}

The {\it Chandra} and the {\it XMM-Newton} 
satellites are currently providing a wealth of x-ray spectra on many 
astronomical objects. There is a serious lack 
of adequate atomic data, particularly in the K-shell energy range, needed for
the interpretation of these spectra.
Spectroscopy in the soft x-ray region (5-45 \AA), including 
K-shell transitions for singly, doubly and triply charged ionic 
forms of atomic elements such as; C, N, O, Ne, S and Si, and the L-shell 
transitions of Fe and Ni, provide a valuable probe of the extreme 
environments in active galactic nuclei (AGN's),
x~ray binary systems, and cataclysmic variables \cite{McLaughlin2001}.  
The goal of the present investigation is to 
provide accurate values for photoionization cross sections,
resonance energies, and  autoionization linewidths resulting from the 
photoabsorption of x~rays near the K-edge of Li-like carbon.  

The synergistic and symbiotic relationship between theoretical
and experimental studies is essential to verify the data produced  from such investigations.
Identification of Auger states from multiply charged ionic states 
of carbon  have been performed experimentally by Schneider and co-workers \cite{Schneider1977}.
Excitation  energies of several autoionizing states in the C$^{3+}$ ion were determined by  
Hofmann \etal \cite{Hofmann1990} by using collisional spectroscopy of fine details in the cross section for electron 
impact ionization of C$^{3+}$ ions.
Jannitti and co-workers \cite{Jannitti1995} were the first to measure photoionization (PI)
cross sections for this Li-like carbon ion over a wide range of energies 
using the dual laser plasma (DLP) technique 
at low spectral resolution compared to the present study. 
Recently, extremely high resolution measurements for
K-shell photoexcitation of singly and doubly charged  
ions of carbon have been carried out within our international collaboration; 
C$\rm ^{+}$ \cite{Schlachter2004} and C$\rm ^{2+}$ \cite{Scully2005}.  
Such studies are important in order to provide 
accurate results for absolute photoionization cross sections, 
resonance energies and autoionization linewidths. These benchmarked results therefore
update existing literature values \cite{Kasstra1993,rm1979,verner1993,verner1995}  
and as such should be used in preference to those that are currently in use 
in the various astrophysical modelling codes such as CLOUDY \cite{Ferland1998,Ferland2003} 
and XSTAR \cite{Kallman2001}.

The present study aims to benchmark theoretical values for PI cross sections, 
resonance energies and lifetimes of autoionizing states of the C$\rm ^{3+}$ ion 
in the vicinity of the K-edge with high-resolution experimental measurements.
This provides confidence in the data that may be used 
 in modelling astrophysical plasmas; e.g.,
in the hot (photoionized and collisionally ionized) gas 
surrounding $\zeta$ Oph \cite{sem1994} where C IV has been observed 
in absorption or for the non-LTE modelling of early B-type stars 
\cite{nieva08}.

Promotion of a K-shell electron in Li-like carbon (C$\rm ^{3+}$) ions 
 to an outer np-valence shell (1s $\rightarrow$ np)  from the ground 
state produces states that can autoionize, forming a C$\rm ^{4+}$ ion
 and an outgoing free electron via the processes;
$$
 h\nu + {\rm C^{3+}(1s^22s)}  \rightarrow  {\rm C^{3+} ~ (1s2snp,~ n=2,3,4 \dots ) }
 $$
 $$
 \downarrow
 $$
 $$
{\rm  C^{4+}~ (1s^2) + e^-.}
$$
The strongest excitation process in the interaction of a photon
with the $\rm 1s^22s~^2S$ ground-state of the Li-like carbon 
ion is the 1s $\rightarrow$ 2p photo-excitation process producing intermediate doubly excited 
C$^{3+}$ ([1s\,(2s\,2p)$^3$P]$^2$P$\rm ^o$) and C$^{3+}$ ([1s\,(2s\,2p)$^1$P]$^2$P$\rm ^o$).  
At higher energy 1s $\rightarrow$ 3p photo-excitation processes  
produce  C$^{3+}$ ([(1s2s)$^3$S~3p] $^2$P$\rm ^o$).
The inner-shell autoionization resonances created by the above processes 
appear in the corresponding photoionization cross sections 
(in the energy region near to the K-edge) on top of a 
continuous background cross section for direct photoionization of the outer 2s electron. 
Indirect and direct photoionization channels can interfere with one another and produce asymmetric (Fano-Beutler)
line profiles. Time-reversed  C$^{3+}$ photoionization processes described by equation (1) constitute 
contributions to dielectronic recombination of C$^{4+}$

The present investigation provides absolute values (experimental and theoretical) for
photoionization cross sections, resonance energies and autoionization linewidths of the
first three intermediate states formed by (1s $\rightarrow$ np) photoexcitation of  C$^{3+}$.
In the ALS experiments, energy scan measurements were taken by stepping the photon
energy through a preset range of values. The energy scan ranges were 
299.8--300.15 eV, 303.29--303.58 eV (at a photon energy spread of $\Delta E = 46$ meV) and 
335.61--337.57 eV (at $\Delta E = 121$ meV) and include the [1s\,(2s\,2p)$^3$P]$^2$P$\rm ^o$,
[1s\,(2s\,2p)$^1$P]$^2$P$\rm ^o$ and [(1s\,2s)$^3$S\,\,3p]$^2$P$\rm ^o$
resonances, respectively. The theoretical photoionization cross section was convoluted with a Gaussian 
of the same full-width at half maximum (FWHM) to simulate the energy resolution of the experiment,
so that direct comparisons may be made with the measurements performed in the various energy regions.
Such a comparison of theoretical  and experimental results 
serves as an indication of the level of accuracy reached by the measurements
and by the theoretical approach \cite{Kjeldsen1999, West2001}.
  
The principle of detailed balancing can be used to compare the present PI 
cross-section measurements with previous experimental and theoretical cross sections for 
the time-inverse photo-recombination (PR) processes. This comparison provides a 
valuable check between entirely different experimental approaches for obtaining atomic cross sections
on absolute scales. 
To benchmark theory and obtain suitable harmony 
with the present high-resolution photoionization 
experimental measurements performed at third-generation 
synchrotron light facilities (such as the Advanced Light Source), 
state-of-the-art theoretical methods are required using 
highly correlated wavefunctions \cite{West2004, Kjeldsen2006}.
Additional theoretical calculations 
are usually required to determine the contribution from ions in 
metastable states which may be present in the parent ion beam 
(which is not an issue in the present case with Li-like ions).
These features have been illustrated vividly from
experimental and theoretical photoionization studies undertaken by  our international collaboration, 
on a number of simple and complex ions.
  All of the experimental work was performed at the Advanced Light Source (ALS), 
in Berkeley, California for a variety of ions, e.g.
He-like Li$\rm ^{+}$ \cite{Scully2006,Scully2007};
Be-like C$\rm ^{2+}$  \cite{Mueller2002,Mueller2003}, 
B$\rm ^{+}$ \cite{Schippers2003},   C$\rm ^{2+}$, N$\rm^{3+}$ and O$\rm^{4+}$ \cite{Mueller2007};
B-like C$\rm ^{+}$  \cite{Schlachter2004};
F-like Ne$\rm ^{+}$  \cite{Covington2002}; 
N-like O$\rm ^{+}$ \cite{Covington2001, Aguilar2003}, F$\rm ^{2+}$ and Ne$\rm ^{3+}$ \cite{Aguilar2005}. 
The majority of these high resolution experimental studies from the ALS have been shown 
 to be in excellent accord with detailed theoretical calculations
performed  using the state-of-the-art R-matrix method \cite{rmat,codes}.

No metastable ions were present in the Li-like C$\rm ^{3+}$ ion beam in the current study.
Photoabsorption experiments in the K-shell region have been performed elsewhere on 
this Li-like carbon ion species at lower resolution than the present experiment 
using the dual laser plasma (DLP) technique \cite{Jannitti1995}.
The DLP measurements have been very useful for obtaining absorption 
spectra over a wide energy range. However their 
interpretation can be complicated due to 
ions being distributed over various charge states in both
the ground and metastable states, and the presence 
of a plasma can affect energy levels.

The layout of this paper is as follows. Section 2 presents a brief outline of the theoretical work. 
Section 3 details the experimental procedure used. Section 4 presents a discussion 
of the results obtained from both the experimental and theoretical methods. 
Finally in section 5 conclusions are drawn from the present investigation.
 
\section{Theory}\label{sec:theory}

Theoretical cross-section calculations for the photoionization of triply 
charged carbon ions are available from the Opacity
Project and can be retrieved from the TOPBASE database \cite{Cunto1993}. 
These cross-section calculations primarily cover the valence region only 
and have been determined in $LS$-coupling.  
Theoretical results from the independent particle model exist 
in the energy region of the K-edge \cite{rm1979,verner1993,verner1995}, but do not 
account for resonance effects that have been observed in the 
dual-laser-plasma (DLP) experimental work of Jannitti and co-workers  or in the 
present study.  Early R-matrix studies for the K-edge region of this ion 
  were limited to $LS$-coupling \cite{McGuiness1997}
with no relativistic or radiation damping effects included in that work.  
No determination of resonance parameters
 were made in that early  R-matrix work \} 
but suitable agreement was found for the background cross section 
with the experimental work  of 
Jannitti and co-workers \cite{Jannitti1995,McGuiness1997}
and the independent particle model \cite{rm1979,verner1993,verner1995}.
Recent studies by Pradhan and co-workers \cite{Nahar2001,Pradhan2001}
for K-shell photoionization cross sections on Li-like complexes have been obtained in 
intermediate coupling (primarily for astrophysical applications)  with identification and 
determination of resonance parameters that will be discussed later in the paper.  

The present investigation extends the earlier $LS$ theoretical work  on the
C$^{3+}$ ion \cite{McGuiness1997}, to include relativistic and radiation damping effects. 
Photoionization cross-section calculations for  C$\rm ^{3+}$ ions were 
performed both in $LS$ and intermediate coupling.  
The intermediate coupling calculations were carried out using the semi-relativistic Breit-Pauli approximation 
which allows for relativistic effects to be included. 
Radiation-damping \cite{damp} effects were also 
included within the confines of the R-matrix approach \cite{rmat,codes} for completeness. 
Relativistic effects need to be included 
when the experimental resolution is such that fine-structure effects 
can be resolved and radiation damping affects the narrow resonances 
present in the PI cross sections.
An appropriate number of C$\rm ^{4+}$ states 
(19 $LS$, 31 $LSJ$ levels) were included in our intermediate coupling calculations. 
 An n=4 basis set of  C$\rm ^{4+}$ orbitals was used
which were constructed using the  atomic-structure code CIV3 \cite{Hibbert1975} to represent the wavefunctions.  
Photoionization cross-section calculations were then performed both in $LS$ and intermediate coupling for the
$\rm 1s^22s~^2S_{1/2}$ initial state of the C$\rm ^{3+}$ ion in order to gauge the importance of 
including relativistic effects.  The photoionization cross-section calculations 
were also performed with and without radiation damping in order to quantify
 this effect in the appropriate PI and PR cross sections.  
It turns out that only the narrow resonances 
in the appropriate cross sections are affected by radiation damping. 

In the calculations the following He-like $LS$ states were retained:
$\rm 1s^2~^1S$,  $\rm 1sns~^{1,3}S$, 
$\rm 1snp~^{1,3}P^{\,\circ}$, $\rm 1snd~^{1,3}D$,  
and $\rm 1snf~^{1,3}F^{\,\circ}$, n $\leq$ 4,
of the C$\rm ^{4+}$ ion core which give rise to 31 $LSJ$ states 
in the intermediate  close-coupling expansions for
the J=1/2 initial scattering symmetry of the Li-like  C$^{3+}$ ion. 
The use of the n=4 pseudo states is to attempt to account for correlations effects and 
the infinite number of states (bound and continuum) 
left out by the truncation of the close-coupling expansion in our work.
For the structure calculations of the 
C$\rm ^{4+}$ ion, all physical orbitals were included up to n=3 in the
configuration-interaction wavefunctions expansions used to describe the states. 

The Hartree-Fock $\rm 1s$ and $\rm 2s$ orbitals of Clementi and Roetti
\cite{Clementi1974} together with the n=3 
 orbitals were determined by energy optimization on the appropriate
spectroscopic state using the atomic structure code CIV3
\cite{Hibbert1975}.  The n=4 correlation (pseudo) orbitals were determined by energy 
optimization on the ground state of this ion. 
All the states of the C$\rm^{4+}$ ion were then represented
by using multi-configuration interaction wave functions.  The Breit-Pauli
$R$-matrix approach was used to calculate the energies 
of the C${\rm ^{4+}}(LSJ)$ states and the subsequent photoionization cross sections. 
A minor shift ($<$  0.1 \%) of the theoretical energies  to experimental values 
\cite{Ralchenko2008} was made so that they would be in agreement with available 
experimental thresholds. 
Photoionization cross sections  out of the 
C$\rm ^{3+}$ (1s$\rm ^2$2s $\rm ^2$S$_{\rm 1/2}$ )
ground-state  were then obtained
for total angular momentum scattering symmetries of 
J = 1/2 and J= 3/2, odd parity, that contribute to the total.

The $R$-Matrix method \cite{rmat,codes,damp} was used to determine 
all the photoionization cross sections for the initial ground state in $LS$ and intermediate-coupling.
  The scattering wavefunctions were generated by
allowing all possible three-electron promotions out of the base $\rm 1s^22s$ configuration of
C$\rm ^{3+}$ into the orbital set employed. 
Scattering calculations were performed with twenty-five
continuum functions and a boundary radius of 8.4 Bohr radii. 
For the $\rm ^2S_{1/2}$ initial state the outer region electron-ion collision 
problem was solved (in the resonance region below and
 between all the thresholds) using a suitably chosen fine
energy mesh of 10$^{-7}$ Rydbergs ($\approx$ 1.36 $\mu$eV) 
to fully resolve all the extremely fine resonance
structure in the appropriate photoionization cross sections. 
The QB technique (applicable to atomic and molecular complexes) 
of Berrington and co-workers \cite{keith1996,keith1998,keith1999} 
was used to determine the resonance parameters and
averaging was performed over final total angular
momentum J values. Finally, in order to compare directly with experiment, 
the theoretical cross section was convoluted with a Gaussian 
function of appropriate width to simulate the energy resolution of the measurement.

\section{Experiment}\label{sec:exp}

The experiment was performed at the ion-photon-beam (IPB)
end-station \cite{Covington2002} of the undulator beamline
10.0.1 at the Advanced Light Source (located at the Lawrence Berkeley
National Laboratory, in Berkeley, California, USA).  The experimental method employed is
similar to that first used by Lyon and co-workers \cite{Lyon1986} and
in our earlier measurements of the K-shell PI cross sections for
the C$\rm ^{2+}$ ion \cite{Scully2005}. 

C$\rm ^{3+}$ ions were generated from CH$_4$ gas inside a compact
all-permanent-magnet electron-cyclotron-resonance (ECR) ion
source \cite{Broetz2001}. Collimated $\rm ^{12}$C$\rm ^{3+}$ 
 ion-beam currents of typically 40 nA were extracted 
by placing the ion source at a positive potential of +6 kV and a dipole 
magnet selected ions of the desired ratio of charge to mass. In addition 
to $^{12}$C$\rm ^{3+}$, the selected ion beam contained other ions with nearly the 
same charge-to-mass ratio, such as $\rm ^{16}$O$\rm ^{4+}$ and $\rm ^{4}He^{+}$. 
The fraction of the measured ion current that was due to $^{12}$C$\rm ^{3+}$ was 
determined to be 87\% from a separate measurement of the 
uncontaminated $^{13}$C$\rm ^{3+}$ ion beam current and applying the 
known $\rm ^{13}$C/$\rm ^{12}$C  natural isotopic abundance ratio of 0.01122. 
This correction to the measured primary ion beam current 
was applied to cross-section measurements.

The ion beam was placed onto the axis of the counter-propagating 
photon beam by applying appropriate voltages to several 
electrostatic ion-beam steering and focusing devices. 
Downstream of the interaction region,  the ion beam was deflected
out of the photon beam direction by a second dipole magnet
that also separated the ionized C$\rm ^{4+}$ product ions from the
C$\rm ^{3+}$ parent ions. The C$\rm ^{4+}$ ions were counted with a
single-particle detector of nearly 100\% efficiency, and the
C$\rm ^{3+}$ ion current was monitored for normalization purposes.
The measured C$\rm ^{4+}$ count rate was only partly due to
photoionization events. It also contained C$\rm ^{4+}$ ions
produced by electron-loss collisions of  C$\rm ^{3+}$ ions 
in the parent beam with residual gas molecules and surfaces. 
For the determination of absolute cross sections
this background was subtracted by time modulation (mechanical chopping) of the
photon beam.

Absolute cross sections were obtained by normalizing the
background-subtracted C$\rm ^{4+}$ count rate to the measured ion
current, to the photon flux, which was measured with a
calibrated photodiode, and to the beam overlap. Beam overlap
measurements were carried out using two commercial
rotating-wire beam-profile monitors and a movable slit
scanner. Due to the considerable effort required for carrying out
reliable absolute cross-section measurements, these were
performed at only a few selected photon energies in the vicinities
of the resonance maxima. The systematic error of the absolute
cross-section determination is estimated to be $\pm$ 20 \% for
the first two resonances. For the third resonance the efficiency of the
photodiode was linearly extrapolated from the lower-energy behaviour
adding 10 to 20 \%  uncertainty to the size 
of the [(1s\,2s)$^3$S\,\,3p]$^2$P$\rm ^o$ peak.

Previous recombination storage-ring measurements 
made at the CRYRING \cite{Mannervik1997,Mannervik1998} and 
the current theoretical calculations guided the high-resolution energy scan 
measurements to be taken by stepping the photon
energy through a preset range of values. 
The scan ranges used  were 299.8--300.15 eV, 303.29--303.58 eV and 335.61--337.57
eV comprising the [1s\,(2s\,2p)$^3$P]$^2$P$\rm ^o$,
[1s\,(2s\,2p)$^1$P]$^2$P$\rm ^o$ and the [(1s\,2s)$^3$S\,\,3p]$^2$P$\rm ^o$
resonances, respectively. Each scan range consisted
of 60 -- 80 data points. The point-to-point step width was
chosen according to the preselected  experimental
energy spread $\Delta E$. The step width was 5 meV in the
first two ranges and 25 meV in the third range.

%+++++++++++++++++++++++++++++++++++++++++++++++++++++++++++++++++++++++++++++
%
%    Figures follow here
%
%    Here is an example of the general form of a figure:
%    Fill in the caption in the braces of the \caption{} command. Put the label
%    that you will use with \ref{} command in the braces of the \label{} command.
%
%+++++++++++++++++++++++++++++++++++++++++++++++++++++++++++++++++++++++++++++

\begin{figure}
%\begin{center}
\includegraphics[width=\textwidth]{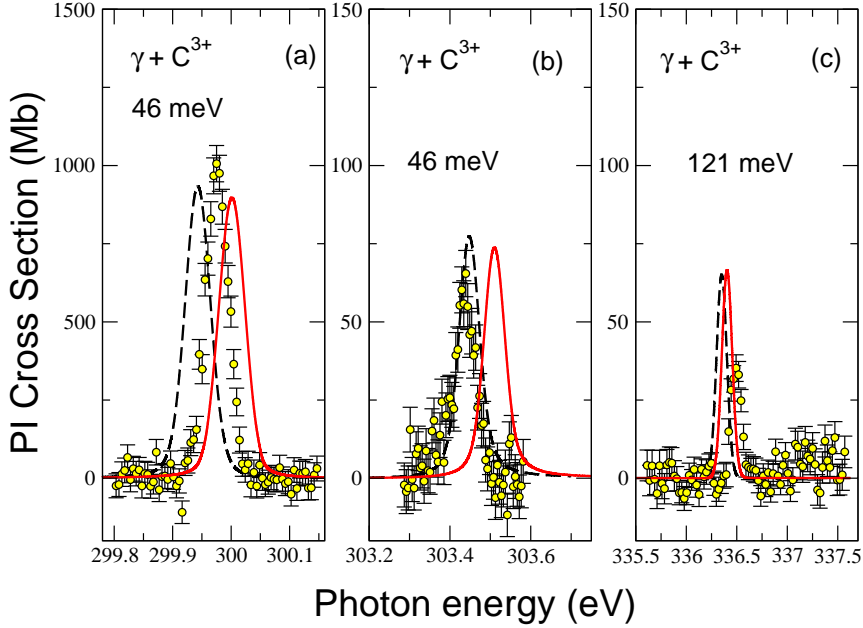}
\caption{\label{fig:C3PIpeaksfinal} (Colour online) Absolute photoionization (PI) cross sections for
            K-shell ionization of Li-like C$\rm ^{3+}$: Experimental
            measurement from the ALS (symbols) and the 
            R-matrix calculations with radiation damping (full lines, Breit-Pauli intermediate coupling,
            dashed lines $LS$ coupling) convoluted with a FWHM
             Gaussian of the appropriate width) for
           (a)  the [1s\,(2s2p)$\rm ^3$P]$\rm ^2$P resonance, (b) the
            [1s\,(2s2p)$\rm ^1$P]$\rm ^2$P resonance, and  (c) the
            [(1s\,2s)$\rm ^3$S\,\,3p]$^2$P resonance. The experimental
            energy spreads obtained from the Voigt fits are 
            (a) $\Delta E = 46$~meV, (b) $\Delta E = 46$ meV
             and (c) $\Delta E = 121$~meV. }
%\end{center}
\end{figure}

The energy scale was calibrated by first applying a Doppler
correction to account for the motion of the C$^{3+}$ ions. The
energy shift resulting from the monochromator calibration was
determined by measurements of the photoions produced 
in photoionization of Ar gas in the vicinity of the Ar 2p$_{3/2}$  edge 
and of CO gas in the vicinity of the C 1s edge. 
The Ar photoabsorption spectrum taken in the energy range 244 to 252 eV 
was compared with the measurements carried out by King \etal \cite{King1977}
who employed electron-energy-loss spectroscopy (EELS) reaching 
unsurpassed precision in this energy range (10 meV at 244.39 eV).  
The present CO photoabsorption spectrum  showing almost fully 
resolved vibrational excitations in CO was compared with the 1s $\to$ $\pi^*$
resonance energies measured by EELS \cite{Sodhi1984} and with the relative resonance energies 
observed in CO photoexcitation \cite{Domke1990} covering an energy range 287 to 306 eV. 
The precision of these measurements is at best 20 meV at 287.40 eV.  
The comparison of the present photoabsorption energies with the literature 
values showed deviations between 0.7 eV near 245 eV and 1.7 eV near 300 eV 
well represented by a linear increase with the monochromator's grating angle. This linear dependence 
was extrapolated all the way to 339 eV yielding a shift of about 2.5 eV. 
The set energies in the experiment were corrected by subtracting the energy shifts 
determined by the linear fit to the observed deviations. This procedure yields an 
uncertainty of $\pm$ 30 meV at 300 eV.  Extrapolation of the calibration range 
to 340 eV, i.e.,  to the very end of the range of the monochromator grating, adds 
another $\pm$ 30 meV if the assumption of a linear dependence is correct. 
An estimate for the uncertainty of  $\pm$ 100 meV at 340 eV appears to be more realistic.

\section{Results and Discussion}\label{sec:res}

\begin{figure}
\includegraphics[width=\textwidth]{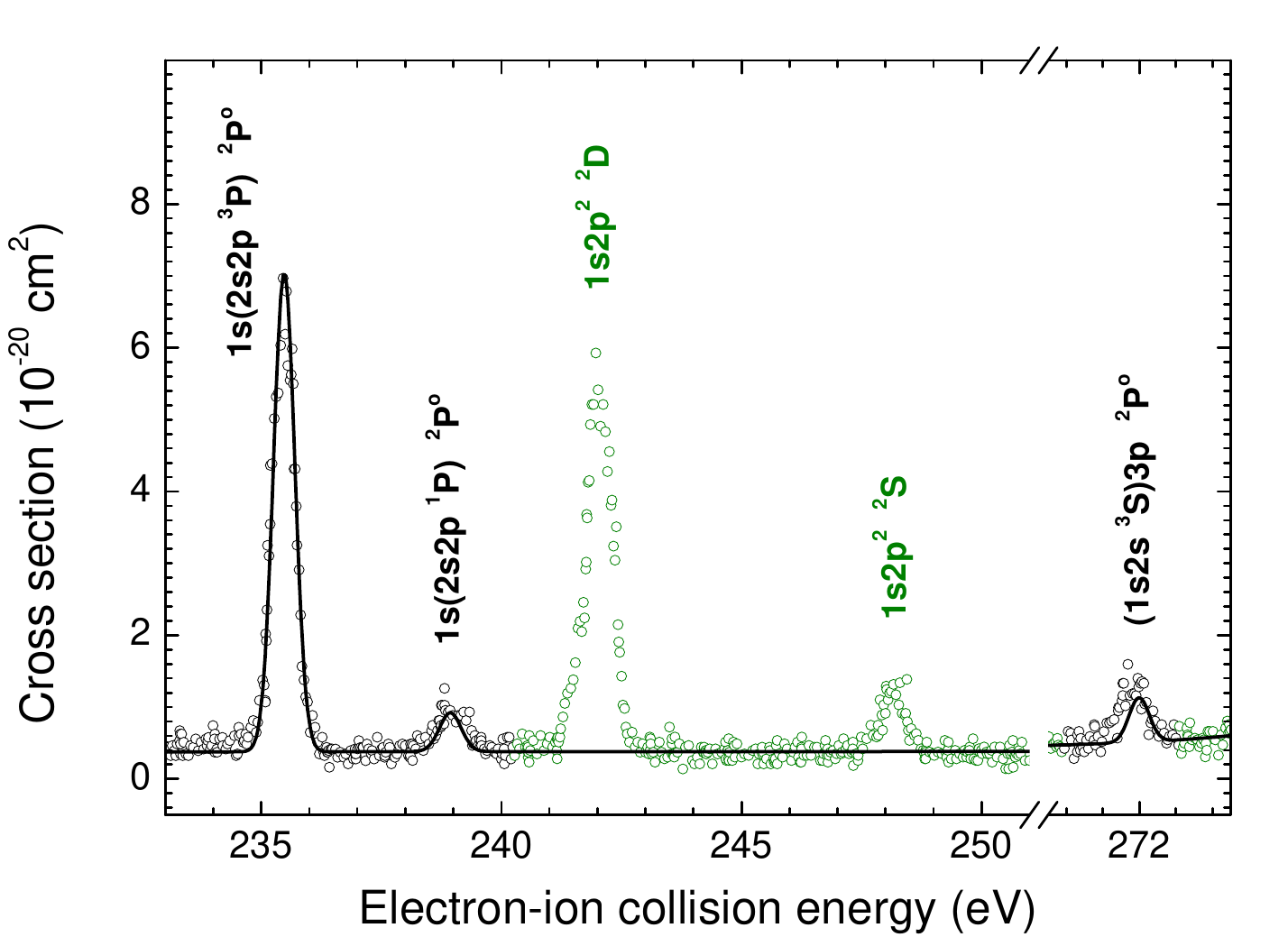}
\caption{\label{fig:C3PIC4PR} (Colour online) Absolute cross sections 
            for the photorecombination (PR) of He-like C$\rm ^{4+}$.  
            Comparison between the experimental C$\rm ^{4+}$ PR results of
            Mannervik et al. \cite{Mannervik1997} (open symbols) from the CRYRING and the
            present experimental C$\rm ^{3+}$ photoionization (PI) results (full line) from the ALS.
	    For comparison purposes the ALS PI cross-sections were converted into PR
            cross-sections (via equation \ref{eq:balance}) and convoluted
            with an appropriate Gaussian to account for the energy spread of the CRYRING PR
            experiment. The [(1s\,2s)$\rm ^3$S\,\,3p]$\rm ^2$P resonance is on the
            tail of a stronger PR-resonance at higher energies (not
            shown). The 1s$\rm ^2$\,2p\,\,$\rm ^2$D and 1s$\rm ^2$\,2p\,\,$\rm ^2$S
            resonances (open green circles) are only observed in the PR experiment since their
            photoexcitation from the ground state of C$^{3+}$ is not dipole allowed. }
\end{figure}

\begin{figure}
\includegraphics[width=\textwidth]{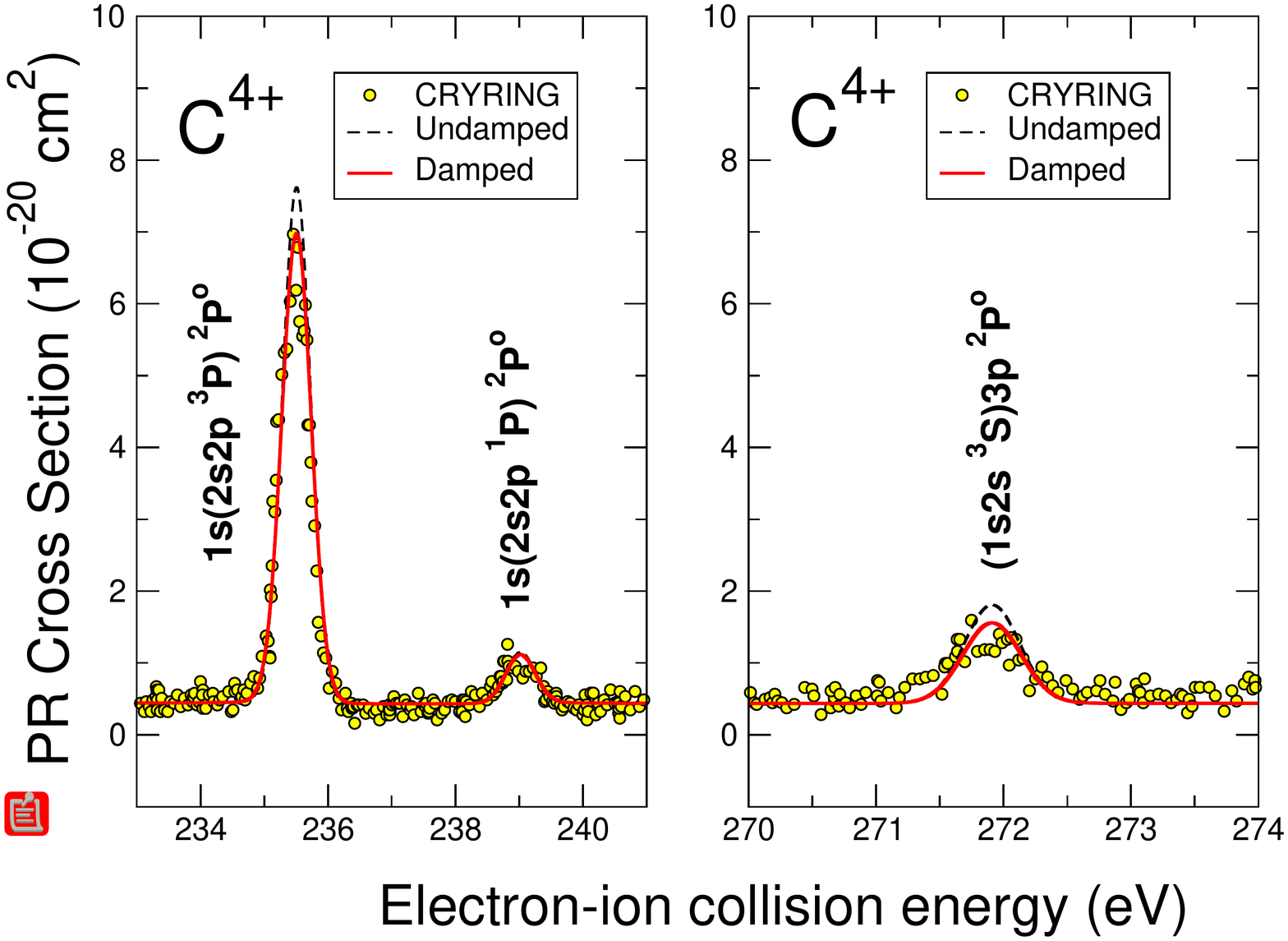}
\caption{\label{fig:C3SIGC4PR} (Colour online) Absolute cross sections 
            for the photorecombination (PR) of He-like C$\rm ^{4+}$. 
            Comparison between the experimental C$\rm ^{4+}$ PR results of
            Mannervik et al. \cite{Mannervik1997} (open symbols) from the CRYRING and the
            present theoretical C$\rm ^{3+}$ results from the 31-state intermediate coupling
            R-matrix method (full line, with radiation damping, 
            dashed line, without radiation damping).  Here again for comparison purposes the
            R-matrix PI cross sections were converted into PR
            cross sections (via equation \ref{eq:balance}) and convoluted
            with an appropriate Gaussian to account for the energy spread of the CRYRING PR
            experiment. }
\end{figure}

The experimental  K-shell photoionization (PI) cross sections for the C$^{3+}$ ion are shown
in figure \ref{fig:C3PIpeaksfinal}. The full line is the result from the R-matrix calculations 
including radiation damping and convolution at the appropriate experimental resolution. 
This convolution masks asymmetric line shapes which become evident when 
zooming in to the cross section range 0 - 1 Mb.
Experimental results for photoionization resonance strengths, level energies and, 
where possible, for Lorentzian widths of the first three Auger states were extracted from 
Voigt line profiles obtained from  nonlinear least-squares fits to the measured data. 
The results from these fits are presented in table \ref{tab:fit1}. 
Table 1 also displays the corresponding results of
the present R-matrix calculations in addition to those
from the most precise experimental and theoretical study to date  of
Mannervik et al. \cite{Mannervik1997}.

%+++++++++++++++++++++++++++++++++++++++++++++++++++++++++++++++++++++++++++++
%
%    Tables follow here
%
%    Here is an example of the general form of a table:
%    Fill in the caption in the braces of the \caption{} command. Put the label
%    that you will use with \ref{} command in the braces of the \label{} command.
%    Insert the column specifiers (l, r, c, d, etc.) in the empty braces of the
%    \begin{tabular}{} command.
%
%+++++++++++++++++++++++++++++++++++++++++++++++++++++++++++++++++++++++++++++

\fulltable{\label{tab:fit1} Comparison of PI resonance energies $E_{\rm ph}^{\rm (res)}$ (in eV), autoionization
         widths $\Gamma$ (meV) and strengths $\overline{\sigma}^{\rm PI}$ (in Mb eV).
         The systematic uncertainty of the present experimental energy scale is 30 meV at 300 eV and 
         an estimated 100 meV at 340 eV. On the basis of their thorough theoretical investigation of 
         resonance energies the authors of reference  \cite{Mannervik1997} felt that the uncertainty of their energy 
         scale was as low as 50 meV, however, this figure could not be derived strictly on 
         experimental grounds.  The systematic uncertainty of the present experimental 
         cross sections is estimated to be 20\% for the two resonances at lower energies 
         and 40\% for the third resonance. The related numbers for the associated PR resonances 
         are not specified in reference \cite{Mannervik1997}. For the comparison, the PR resonance strengths from 
         reference \cite{Mannervik1997} were converted by employing equation  (1). 
         The relative energies $E_{\rm ph}^{\rm (res)}$(1) and  $E_{\rm ph}^{\rm (res)}$(2) of the first two resonances were determined 
         in the experiment with a precision of 1 to 2 meV, from which a more accurate number for the energy splitting 
         $\Delta E_{\rm res}$ (in eV) =  $E_{\rm ph}^{\rm (res)}$(2) - $E_{\rm ph}^{\rm (res)}$(1) can be inferred.}
         
\begin{tabular}{@{}cccccc}
\br
 Resonance                                          &                                                           & ALS                             & R-matrix     	                 & CRYRING 	         & SPM\\ 
 							&						   &					&					&				&(MCDF)\\
                                                                &                                                           &(Expt)                            &(Theory)    	                 &   (Expt)                       &(Theory)\\
 \ns
 \mr
 \lineup  
$$[1s(2s2p)$^3$P]$^2$P$\rm ^o$ & $E_{\rm ph}^{\rm (res)}$ (1) & 299.98 $\pm$ 0.03    & 299.99$^{\ddagger}$        	& 299.98 $\pm$ 0.05    & 299.99$^{\ast}$\\
  &                                                                                                                &                                       & 299.94$^{\dagger}$          	&                                       &                 \\
  \\
 & $\Gamma$                                                                                            &  \ \-                                 & 9.5$^{\ddagger}$             	& \ \-                                & \, 3.88$^{\ast}$   \\
 &                                                                                                                 &                                       & 9.5$^{\dagger}$             	 	&                                     & 9.5$^{\S}$ \\
 \\
 & $\overline{\sigma}^{\rm PI}$                                                              & \,53 $\pm$ 2                &  53.3$^{\ddagger}$            	& \,52.6 $\pm$ 0.8       & \ \-         \\
 &                                                                                                                 &                                       &  54.4$^{\dagger}$           		&                                     &                 \\
\\
$$[1s(2s2p)$^1$P]$^2$P$\rm ^o$ & $E_{\rm ph}^{\rm (res)}$ (2) & 303.44 $\pm$ 0.03    & 303.50$^{\ddagger}$          	& 303.48 $\pm$ 0.05   & 303.46$^{\ast}$   \\
  &                                                                                                                &                                       & 303.43$^{\dagger}$          	&                                      &                  \\
  \\
 & $\Gamma$                                                                                            & \,27 $\pm$ 5                & 26.0$^{\ddagger}$             	& \ \-                                &\, 39.91$^{\ast}$      \\
&                                                                                                                  &                                       & 25.6$^{\dagger}$              	 &                                     &25.4$^{\S}$\\
 \\ 
 & $\overline{\sigma}^{\rm PI}$                                                              & \,\,4.5 $\pm$ 0.7          &  5.6$^{\ddagger}$              	&\,\,6.5 $\pm$ 0.4         & \ \-          \\
 &                                                                                                                 &                                        & 5.8$^{\dagger}$            		&                                      &                \\
\\
$$[(1s\,2s)$^3$S\,\,3p]$^2$P$\rm ^o$ &$E_{\rm ph}^{\rm (res)}$(3) & 336.50 $\pm$ 0.10     & 336.39$^{\ddagger}$         & 336.36 $\pm$ 0.05   & 336.39$^{\ast}$   \\
   &                                                                                                               &                                        & 336.33$^{\dagger}$        	 &                                      &                 \\
   \\
 & $\Gamma$                                                                                            & \ \-                                  & 0.13$^{\ddagger}$          	&  \ \-                                  & 0.55$^{\ast}$     \\
 &                                                                                                                 &                                       & 0.13$^{\dagger}$          		&                                         &0.13$^{\S}$\\
 \\
 & $\overline{\sigma}^{\rm PI}$                                                              & \,\,4.5 $\pm$ 0.9          & 8.6$^{\ddagger}$                 	& \,\,7.0 $\pm$ 0.4          &  \ \-            \\
 &                                                                                                                 &                                        & 8.5$^{\dagger}$                  	&                                       &                  \\
 \\
Energy splitting & $\Delta E_{\rm res}$                                                        & \,\,3.461$\pm$ 0.004   & 3.509$^{\ddagger}$          	& \,\,3.50                          &  3.465$^{\ast}$ \\
&                                                                                                                  &                                          &  3.499$^{\dagger}$           	&                                       &                   \\
 \br
\end{tabular}
\\
$^{\ddagger}$Breit-Pauli semi-relativistic intermediate coupling R-matrix (31-state).\\
$^{\dagger}$Non-relativistic $LS$ coupling R-matrix (19-state).\\
$^{\ast}$Saddle-Point-Method \cite{Mannervik1997}.\\
$^{\S}$MCDF method \cite{Chen1986}.\\
\\
\endfulltable

Mannervik and collaborators performed an electron-ion recombination experiment with 
a cooled ion beam at a heavy-ion storage-ring \cite{Mannervik1997,Mannervik1998}, 
and observed doubly excited C$^{3+}$ resonance states in photorecombination (PR)
of C$^{4+}$ ions (figure \ref{fig:C3PIC4PR}). Photorecombination of C$^{4+}$
is the time-reversed process of C$^{3+}$ PI. The corresponding
cross sections $\sigma^{\mathrm{PR}}$ and
$\sigma^{\mathrm{PI}}$ can be compared on a state-to-state
level \cite{Mueller2002,Schippers2002,Schippers2004} by
employing the principle of detailed balance:
\begin{equation}\label{eq:balance}
    \sigma^{\mathrm{PR}}_{f \to i} = 
    \sigma^{\mathrm{PI}}_{i \to f} \frac{g_i}{g_f}\frac{E_\mathrm{ph}^2}{2m_ec^2E_e}
\end{equation}
Here $g_i = 2$ is the statistical weight of the
C$\rm ^{3+}$(1s$\rm ^2$\,2s\,\,$\rm ^2$S$\rm{_{1/2}}$) ground state (labelled
$i$), $g_f = 1$ is the statistical weight of the
C$\rm ^{4+}$(1s$\rm ^2$\,\,$\rm ^1$S$\rm{_0}$) ground state (labelled $f$). The
PR and PI energy scales (denoted as $E_\mathrm{ph}$ and $E_e$,
respectively) differ by the C$\rm ^{3+}$ $\rm 2s$-ionization energy
$I_i = 64.49390$~eV \cite{Ralchenko2008}, i.\,e.\ $E_e = E_\mathrm{ph} - I_i$.

Photorecombination generally leads to a multitude of final states. In the case
of the 1s\,2s\,np resonances, however, it may be assumed
that only the radiative transition to the 1s$\rm ^2$\,2s ground
state leads to a final state that is stable against
autoionization. Therefore, the application of equation
\ref{eq:balance} facilitates a direct comparison of the
resonance parameters obtained from the present PI experiment
with the ones from the PR experiment. 

\fulltable{\label{tab:fit2} Autoionization  ($\Gamma_a$) and radiative rates ($\Gamma_r$)  with branching ratios $\eta$ (\%)
                                          for the 1s2$\ell$2$\ell^{\prime}$ resonance states of the C$^{3+}$ ion. The present results
                                          are determined in $LS$ coupling.  The MCDF calculations taken from reference \cite{Chen1986}  were 
                                           averaged over fine-structure levels.  The theoretical results from the Saddle-Point-Method (SPM) were taken 
                                           from reference \cite{Mannervik1997}.   The numbers in the square brackets 
                                          denotes the power of 10 by which the preceding term is to be multiplied. }

\begin{tabular}{@{}ccccc}
\br
C$^{3+}$(1s2$\ell$2$\ell^{\prime}$)          & Present                                                  	&  Present   	               	& Other			& Other		\\
States							 &								&					& Methods		& Methods	\\
                                                                   	 &   $\Gamma_a$                                           &  $\Gamma_r$			&   $\Gamma_a$      &   $\Gamma_r$     \\
                                                                    	 &   (meV, s$^{-1}$)                                       &  (meV, s$^{-1}$)		&   (meV, s$^{-1}$)    &  (meV, s$^{-1}$)      \\
 \ns
 \mr
 \lineup
$$[1s(2s2p)$^3$P]$^2$P$\rm ^o$  		&								&					&							&                             \\   
							 	&    9.5,  1.44[13]					& 0.430, 6.54[11] 		&9.50,   1.43[13]$^{\dagger}$        	&0.322, 4.89[11]$^{\dagger}$  \\
								&                                                                        &                                            & 3.88,   0.59[13]$^{\ddagger}$		&0.442,  6.72[11]$^{\ddagger}$ \\
 \\                                                                   
$$[1s(2s2p)$^1$P]$^2$P$\rm ^o$   		&								&					&							&			  \\	
								& 25.6, 3.89[13]					&0.057, 8.70[10] 		&25.40,  3.86[13]$^{\dagger}$       	&0.063, 9.57[10]$^{\dagger}$ \\ 
								&                                                                        &               				&39.91,   6.07[13]$^{\ddagger}$	&0.049, 7.41[10]$^{\ddagger}$  \\
   
\\
$$[(1s\,2s)$^3$S\,\,3p]$^2$P$\rm ^o$ 	&								&					&				&                             \\ 
								&0.133, 2.02[11]					&0.087, 1.33[11] 		&0.127,  1.93[11]$^{\dagger}$		&0.082, 1.25[11]$^{\dagger}$   \\
								&								&					&0.550,  8.36[11]$^{\ddagger}$	&0.086, 1.30[11]$^{\ddagger}$\\ 
\\
									&		&		           				& $\eta$ (\%)				                              &		\\
\\
									&		&  $$[1s(2s2p)$^3$P]$^2$P$\rm ^o$ & $$[1s(2s2p)$^1$P]$^2$P$\rm ^o$ & $$[(1s\,2s)$^3$S\,\,3p]$^2$P$\rm ^o$ \\
\\
$\Gamma_r$/($\Gamma_a$ + $\Gamma_r$)   & Radiative			&  \,\,4.32$^{\S}$				&0.22$^{\S}$					& 36.67$^{\S}$	\\
									&					&  \,\,3.28$^{\dagger}$			&0.25$^{\dagger}$				& 39.23$^{\dagger}$	\\
									&					& 10.20$^{\ddagger}$			&0.12$^{\ddagger}$		        		&13.50$^{\ddagger}$ \\
\\
$\Gamma_a$/($\Gamma_a$ + $\Gamma_r$)  & Autoionization		&95.68$^{\S}$				       	&99.78$^{\S}$					&63.33$^{\S}$	\\
									&					&96.72$^{\dagger}$				&99.75$^{\dagger}$				&60.77$^{\dagger}$ \\
									&					&89.80$^{\ddagger}$			&99.88$^{\ddagger}$			&86.50$^{\ddagger}$  \\		 
 \br
\end{tabular}
\\
 $^{\S}$Present $LS$ coupling work.\\
 $^{\dagger}$MCDF method \cite{Chen1986}.\\
 $^{\ddagger}$Saddle-Point-Method (SPM)  \cite{Mannervik1997}.\\
\\
\endfulltable

For the comparison presented in figure
\ref{fig:C3PIC4PR} the present experimental PI cross sections
were converted into PR cross sections (via equation
\ref{eq:balance})
and convoluted with a Gaussian to account for the experimental
energy spread $\Delta E$ of the PR experiment. The latter was
determined from a fit of the converted PI cross section to the
experimental PR cross section (full line in figure
\ref{fig:C3PIC4PR}). The fit delivered 
$\Delta E = 520 \pm 60$~meV, i.\,e., an 
order of magnitude larger than in the
present PI experiment. 

In the comparison of the converted PI results with the PR data, 
an overall cross section scale factor was allowed to vary in order 
to match the converted PI and the experimental PR cross-section scales. 
This factor was determined to be 0.93 $\pm$ 0.16, i.e., both 
cross-section scales agree with one another within an uncertainty of 16\%. 
The actual observed deviation from unity is within the combined systematic 
uncertainties of both experiments. In reference \cite{Mannervik1997} no systematic uncertainty 
for the cross section scale is provided, however, error bars of 20\% are 
typical for storage ring recombination experiments and have been assumed here as well.

A closer inspection of the individual resonance strengths (table \ref{tab:fit1}) highlights 
the agreement between the cross-section scales of the 
PI and PR experiments, especially for the strongest 
[1s(2s2p)$\rm ^3$P]$\rm ^2$P$\rm ^o$ resonance. The error bars for the integrated 
resonance strengths are statistical only. The ratios between the
individual strengths from reference \cite{Mannervik1997} and
the present ones are 0.992, 1.44 and 1.56 in the order of table \ref{tab:fit1}. Obviously
there is less agreement for the weaker [1s(2s2p)$^1$P]$^2$P$\rm ^o$
and [(1s\,2s)$\rm ^3$S\,\,3p]$\rm ^2$P$\rm ^o$ resonances.  
This might partly be attributed to the fact that in addition to the 1s$\rm ^2$\,2s\,$\rm ^2$S
ground state there are more final states available for dielectronic recombination (DR) via
the [1s(2s2p)$\rm ^1$P]$\rm ^2$P$\rm ^o$ and [(1s\,2s)$\rm ^3$S\,\,3p]$\rm ^2$P$\rm ^o$
resonances.  

Both experiments agree with each other within the systematic uncertainties 
(30 meV for the present PI experiment and 50 meV for the PR experiment) for 
the [1s(2s2p)$\rm ^3$P]$\rm ^2$P$\rm ^o$ and [1s(2s2p)$\rm ^1$P]$\rm ^2$P$\rm ^o$ resonance energies.
The present experimental resonance energies are lower by 4 meV and 44 meV, 
respectively, than those from reference \cite{Mannervik1997}. The agreement with the theoretical 
results from reference \cite{Mannervik1997}  is excellent with differences of only 14 meV and 19 meV, respectively. 
The present intermediate coupling R-matrix calculations yield [1s(2s2p)$\rm ^3$P]$\rm ^2$P$\rm ^o$ 
 and  [1s(2s2p)$\rm ^1$P]$\rm ^2$P$\rm ^o$ resonance 
energies of 299.991 eV and 303.500 eV which are respectively 15 meV and 64 meV 
higher than our experimental values; 11 meV and 20 meV, compared to the CRYRING experiment.  
It is interesting to compare the energy difference $\Delta E_{res}$ =  $E_{\rm ph}^{\rm (res)}$(2) - $E_{\rm ph}^{\rm (res)}$(1) 
found in the ALS experiment (see entry  $\Delta E_{\rm res}$ in table 1).  The SPM approach is within the experimental uncertainty of 
the ALS experiment. The present intermediate coupling R-matrix calculations differ by only 10 meV 
from the $LS$ R-matrix result, which again differs by 38 meV from  the ALS result.
We note that from earlier dielectronic recombination (DR) measurements taken at 2 eV resolution at the TSR 
storage ring in Heidelberg, Germany  \cite{Kilgus1993}  less accurate results (compared to the 
present experimental values of 299.976 $\pm$ 0.03 eV and 303.436 $\pm$ 0.03 eV) 
of 298.794 $\pm$ 0.1 eV and 302.293 $\pm$ 0.3 eV, 
respectively, were obtained, as were theoretical predictions of 299.3 eV and 302.28 eV 
made by Pradhan and co-workers \cite{Nahar2001,Pradhan2001} using the Breit-Pauli R-matrix method for the 
energies of these same resonances.  
The theoretical results of the Saddle-Point-Method (SPM) were taken from reference \cite{Mannervik1997}
and the MCDF values are from reference \cite{Chen1986} averaged over fine-structure levels. 
From figure 1 and table 1 we see that the non-relativistic R-matrix results
for the energies of all three resonances  yield consistently lower values compared to those from
more sophisticated theoretical approaches (which include relativistic effects) and with experiment. 
The autoionization linewidths and resonance strengths are of
similar magnitude to the intermediate coupling R-matrix results. 
We note that the energy position of  the [1s(2s2p)$\rm ^1$P]$^2$P$\rm ^o$ broad resonance 
from the $LS$ coupling results is in better agreement with the ALS experiment.

The present experimental value  of 336.492 $\pm$ 0.1 eV for the [(1s\,2s)$^3$S\,\,3p]$^2$P$\rm ^o$ 
 resonance  position has a deviation of 132 meV between the  
 ALS work and the value of 336.36 $\pm$ 0.05 eV from that of the CRYRING \cite{Mannervik1997}. 
This we attribute to a deficiency of the present experimental energy calibration. 
The resonance energy of 336.4 eV is well outside the energy range where calibration 
lines were measured with an uncertainty of $\pm$ 30 meV (section \ref{sec:exp}). The extrapolation 
of the present calibration to energies well outside the investigated range introduces 
additional uncertainties that were estimated to result in a possible 0.1 eV error.  
The present intermediate coupling R-matrix value of 336.393 eV for the energy position of this 
resonance lies 33 meV above the experimental value from the CRYRING and 
thus still within its 50 meV experimental uncertainty.  

Table 2 presents our $LS$ results for the autoionization ($\Gamma_a$), radiative ($\Gamma_r$) rates 
and branching ratios  ($\eta$) for the above three 1s2$\ell$2$\ell^{\prime}$ resonance states of C$^{3+}$ together with results from   
the multi-configuration-Dirac-Fock (MCDF) approach \cite{Chen1986} and the saddle point method \cite{Mannervik1997}. 
We note from table 2 that our $LS$ results for these quantities show better accord with the MCDF results \cite{Chen1986} than with those from the 
saddle-point-method \cite{Mannervik1997} and that inclusion of radiation damping in the R-matrix calculations will 
  only affect the two narrow resonances observed in both the ALS and CRYRING spectra.  
This is illustrated clearly in figure 3 where it is seen that radiation damping only affects the theoretical R-matrix results 
for the narrow resonances present in the PR cross sections.  Finally, the good agreement  of the 
present PI converted intermediate coupling R-matrix results with the PR experimental data (figure \ref{fig:C3SIGC4PR}) obtained at 
the CRYRING  by Mannervik  and co-workers \cite{Mannervik1997}, provides  further confidence in our present work. 

\section{Conclusion}
State-of-the-art theoretical and experimental methods were used to
study the photoionization of C$^{3+}$ ions in the energy  region near to the K-edge. 
Overall, agreement is found between the present theoretical and 
experimental results both on the photon-energy scale and on the
absolute PI cross-section scale for this prototype Li-like system. 

The strength of the present study is in its excellent
experimental resolving power coupled with state-of-the-art theoretical predictions. 
The experimental energy resolution of 46~meV in the present work
made possible a determination of the autoionization linewidth of the
[1s(2s2p)$^1$P]$^2$P$\rm ^o$ resonance. 
The Voigt line-profile fit  for this resonance yielded a value for the 
autoionization linewidth of $27 \pm 5$~meV  which is in good agreement with the present
theoretical prediction of  26 meV (table \ref{tab:fit1}) and 
nearly  50\% smaller than the theoretical result of Mannervik \etal  \cite{Mannervik1997}.  The 
energy resolution and calibration of the present PI experiment also made possible to determine the 
energy difference $\Delta E_{\rm res}$ between the C$^{3+}$ ([1s\,(2s\,2p)$^1$P]$^2$P$\rm ^o$) 
and C$^{3+}$ ([1s\,(2s\,2p)$^1$P]$^2$P$\rm ^o$) resonances with an uncertainty of only 4 meV. 
The difference $\Delta E_{\rm res}$ = 3.461 $\pm$ 0.004 eV is in agreement with the theoretical result 
of the saddle point method used by Mannervik \etal  \cite{Mannervik1997}.

The principle of detailed balance was used to compare the present PI 
cross-section measurements with previous experimental and theoretical cross sections for 
the time-inverse photo-recombination (PR) processes. This provides a
valuable check between entirely different approaches for 
obtaining atomic cross sections on absolute scales and 
gives confidence in the accuracy of the results.

\ack
We acknowledge support by Deutsche Forschungsgemeinschaft under project number Mu 1068/10, 
by the US Department of Energy (DOE) under contract DE-AC03-76SF-00098 and grant DE-FG02-03ER15424, 
 through PAPIIT IN108009 UNAM, and through NATO Collaborative Linkage grant 976362. 
We thank Sven Mannervik for providing the numerical 
data of the Stockholm C$^{4+}$ recombination measurements.
B M McLaughlin acknowledges support by the US
National Science Foundation through a grant to ITAMP
at the Harvard-Smithsonian Center for Astrophysics.
%+++++++++++++++++++++++++++++++++++++++++++++++++++++++++++++++++++++++++++++
%
%   Reference section now follows
%
%   Delete or change fake bibitem. delete next three
%   lines and directly read in your .bbl file if you use bibtex.
%
%+++++++++++++++++++++++++++++++++++++++++++++++++++++++++++++++++++++++++++++
%
\clearpage

\bibliographystyle{iopart-num}

\bibliography{c3plus}

\end{document}